\documentclass[twocolumn,english,superscriptaddress,citeautoscript,showpacs,preprintnumbers,amsmath,amssymb,prb,floatfix,footinbib]{revtex4}
\usepackage[utf8]{inputenc}
\usepackage[scaled=2]{helvet}
\usepackage[T1]{fontenc}
\usepackage[utf8]{luainputenc}
\usepackage{multirow}
\usepackage{textcomp}
\usepackage{amsmath}
\usepackage{amssymb}
\usepackage{graphicx}
\usepackage{subscript}
\usepackage{epstopdf}
\makeatletter

\providecommand{\tabularnewline}{\\}

\@ifundefined{textcolor}{}
{%
 \definecolor{BLACK}{gray}{0}
 \definecolor{WHITE}{gray}{1}
 \definecolor{RED}{rgb}{1,0,0}
 \definecolor{GREEN}{rgb}{0,1,0}
 \definecolor{BLUE}{rgb}{0,0,1}
 \definecolor{CYAN}{cmyk}{1,0,0,0}
 \definecolor{MAGENTA}{cmyk}{0,1,0,0}
 \definecolor{YELLOW}{cmyk}{0,0,1,0}
}

\usepackage[scaled=2]{helvet}\usepackage{amstext}

\makeatletter%%%%%%%%%%%%%%%%%%%%%%%%%%%%%% Textclass specific LaTeX commands.
\@ifundefined{textcolor}{}{\definecolor{BLACK}{gray}{0}\definecolor{WHITE}{gray}{1}\definecolor{RED}{rgb}{1,0,0}\definecolor{GREEN}{rgb}{0,1,0}\definecolor{BLUE}{rgb}{0,0,1}\definecolor{CYAN}{cmyk}{1,0,0,0}\definecolor{MAGENTA}{cmyk}{0,1,0,0}\definecolor{YELLOW}{cmyk}{0,0,1,0}}

\usepackage{dcolumn}% Align table columns on decimal point
\usepackage{bm}% bold math
\usepackage{times}% font that prx use
\usepackage{hyperref}
\hypersetup{colorlinks=true,linkcolor=red,citecolor=blue}
\makeatother

\usepackage{babel}

\makeatother

\usepackage{babel}
\begin{document}
\title{Spin reorientation in FeCrAs revealed by single-crystal neutron diffraction}

\author{W. T. Jin}
\email{wjin@physics.utoronto.ca}
\affiliation{Department of Physics, University of Toronto, Toronto, Ontario, M5S 1A7, Canada}

\author{M. Meven}
\affiliation{RWTH Aachen University, Institut für Kristallographie, D-52056 Aachen, Germany}
\affiliation{Jülich Centre for Neutron Science JCNS at Heinz Maier-Leibnitz Zentrum (MLZ), Forschungszentrum Jülich GmbH, Lichtenbergstraße 1, D-85747 Garching, Germany}

\author{H. Deng}
\affiliation{RWTH Aachen University, Institut für Kristallographie, D-52056 Aachen, Germany}
\affiliation{Jülich Centre for Neutron Science JCNS at Heinz Maier-Leibnitz Zentrum (MLZ), Forschungszentrum Jülich GmbH, Lichtenbergstraße 1, D-85747 Garching, Germany}

\author{Y. Su}
\affiliation{Jülich Centre for Neutron Science JCNS at Heinz Maier-Leibnitz Zentrum (MLZ), Forschungszentrum Jülich GmbH, Lichtenbergstraße 1, D-85747 Garching, Germany}

\author{W. Wu}
\affiliation{Department of Physics, University of Toronto, Toronto, Ontario, M5S 1A7, Canada}

\author{S. R. Julian}
\affiliation{Department of Physics, University of Toronto, Toronto, Ontario, M5S 1A7, Canada}

\author{Young-June Kim}
\email{yjkim@physics.utoronto.ca}
\affiliation{Department of Physics, University of Toronto, Toronto, Ontario, M5S 1A7, Canada}

\begin{abstract}
The magnetic structure of the ``nonmetallic metal'' FeCrAs, a compound with the characters of both metals and insulators, was examined as a function of temperature using single-crystal neutron diffraction. The magnetic propagation vector was found to be $\mathit{k}$ = (1/3,
1/3, 0), and the magnetic reflections disppeared above $\mathit{T_{N}}$ = 116(1) K. In the ground state, the Cr sublattice shows an  in-plane spiral antiferromagnetic order. The moment sizes of the Cr ions were found to be small, due to strong magnetic frustration in the distorted Kagome lattice or the itinerant nature of the Cr magnetism, and vary between 0.8 and 1.4 $\mu_{B}$ on different sites as expected for a spin-density-wave (SDW) type order. The upper limit of the moment on the Fe sublattice is estimated to be less than 0.1 $\mu_{B}$. With increasing temperature up to 95 K, the Cr moments cant out of the $\mathit{ab}$ plane gradually, with the in-plane components being suppressed and the out-of-plane components increasing in contrast. This spin-reorientation of Cr moments can explain the dip in the $\mathit{c}$-direction magnetic susceptibility and the kink in the magnetic order parameter at $\mathit{T_{O}}$ \textasciitilde{} 100 K, a second magnetic transition which was unexplained before. We have also discussed the similarity between FeCrAs and the model itinerant magnet Cr, which exhibits spin-flip transitions and SDW-type antiferromagnetism.

\end{abstract}
\date{Oct. 31, 2019}

\maketitle
%75.25.+z Spin arrangements in magnetically ordered materials(including neutron and spin-polarized electron studies,synchrotron-source X-rayscattering, etc.)
%75.40.Cx Static properties (order parameter, static susceptibility, heat capacities, critical exponents, etc.)
%75.50.Ee Antiferromagnetics

\section{Introduction}

Identifying the nature of magnetism, itinerant or localized, remains a difficult challenge in condensed matter physics. Although purely localized moments were found in many magnetic insulators, in metallic compounds including doped cuprates,\cite{Keimer_15} iron-based superconductors,\cite{Dai_15} and heavy-fermion systems,\cite{Stewart_01} itinerant moments were often found to coexist with the localized moments, which impedes the thorough understanding of the role of magnetism in novel phenomena such as unconventional superconductivity. Recently, one interesting compound with the characters of both itinerant and localized magnetism, FeCrAs, has attracted much attention due to its intriguing properties as a \textquotedblleft nonmetallic metal\textquotedblright .\cite{Wu_09,Akrap_14}
On one hand, specific heat measurement suggests a metallic, highly enhanced Fermi-liquid behavior with a large Sommerfeld coefficient ($\gamma$ $\sim$ 30 mJ/mol K$^{2}$). On the other hand, its electrical resistivity $\rho$(T) deviates strongly from the $\mathit{T^{2}}$
behavior expected for a Fermi liquid and shows a negative slope over a huge tempearature range, from nearly 900 K down to 80 mK.

FeCrAs has a hexagonal Fe$_{2}$P structure (with the space group $\mathit{P}\bar{6}2m$).\cite{Fruchart_82} The Fe ions form a triangular lattice of trimers, while the Cr ions form a distorted Kagome framework within the basal plane. These planes of Fe trimers and Cr Kagome framework stack alternately along the $\mathit{c}$-axis, with the As ions interspersed throughout both layers and in between. Despite strong geometric frustration of the Cr sublattice, FeCrAs was still found to order antiferromagnetically below $\mathit{T_{N}}$ = 125 K.\cite{Wu_09} The magnetic propagation vector was determined to be $\mathit{k}$ = (1/3, 1/3, 0) via neutron diffraction measurements.\cite{Swainson_10,Plumb_08} Based on the base-temperature diffraction pattern from powder samples, the magnetic intensities were found to fit well with a non-collinear antiferromagnetic model in which the moments only reside on the Cr sublattice.\cite{Swainson_10} In the ground state, the Cr moments at different crystallographic sites vary between 0.6 and 2.7 $\mu_{B}$ in the form of a spin density
wave, but the moments on the Fe sublattice were found to be negligibly small. The absence of magnetism in the Fe sublattice is consistent with the prediction from electronic structure calculations.\cite{Ishida_96} The issue of Fe site magnetism has been examined by various experimental methods covering a wide range of time scales. Slow probes such as M\"ossbauer spectroscopy\cite{Footnote} and muon spin relaxation,\cite{Huddart_19} as well as fast probes such as x-ray emission spectroscopy\cite{Gretarsson_11} and resonant magnetic scattering\cite{Huddart_19} all reported that Fe moments are either absent or negligibly small. The nonmagnetic Fe sublattice, however, was argued to play an important role in stabilizing the magnetic ordering through the Fe-Cr coupling, and the strong charge fluctuations in the Fe sublattice due to the proximity to a metal-insulator transition were proposed to be responsible for the unusual transport properties of FeCrAs.\cite{Rau_11} Recently, inelastic neutron scattering (INS) experiment on powder samples of FeCrAs has revealed steeply dispersing, gapless spin wave excitations persisting up to at least 80 meV.\cite{Plumb_08} Such a large energy scale indicates an underlying magnetic energy scale that is significantly larger than that estimated from the local moment model, and resembles the spin-wave-like excitations
in classic itinerant magnets such as the Cr metal.\cite{Fawcett_88} The presence of these stiff high-energy spin fluctuations was suggested as a possibly being connected its ``nonmetallic metal'' properties.

To better understand the intriguing magnetism of FeCrAs, complete knowledge about its ground-state magnetic structure, as well as how it envolves with temperature, is undoubtedly of great importance. In fact, one anomaly in the temperature dependence of its magnetic
susceptibility was not paid much attention. The dc magnetic susceptibility ($\chi$), as shown in Ref. \onlinecite{Wu_09} and Fig. 2 of the current paper, displays a clear dip in $\chi_{c}$ and a bump in $\chi_{ab}$ at the temperature not far below $T_{N}$, but the nature of such an anomaly was never clarified in previous studies. The only available neutron diffraction probe so far to directly determine the magnetic structure of FeCrAs, was performed at 2.8 K on powder samples using thermal neutrons (wavelength $\lambda$ = 2.37 Å).\cite{Swainson_10} Availability of a high-quality single-crystal sample motivated us to revisit the magnetic ground state of FeCrAs using epithermal neutrons to access more magnetic reflections, and to study the evolution of the magnetic structure with temperature in detail.

Here we present the results of our single-crystal neutron diffraction measurements on FeCrAs. We found that the Cr sublattice shows an in-plane spiral antiferromagnetic order. With increasing temperature up to 95 K, the Cr moments cant out of the $\mathit{ab}$ plane gradually, with the in-plane components being suppressed and the out-of-plane components increasing in contrast. Such a spin-reorientation provides a natural explanation of the dip in $\chi_{c}$ around 100 K. A magnetic phase diagram of FeCrAs consisting of three distinct phase regions is proposed.

\section{Experimental Details}

Single crystals of FeCrAs were grown by melting stoichiometric quantities of high-purity Fe, Cr and As following the recipe in Ref. \onlinecite{Katsuraki_66}, and well characterized by transport and thermodynamic measurements in Ref. \onlinecite{Wu_09}. A 79 mg single crystal from the same batch was selected for neutron diffraction measurements, which was performed on the hot-neutron four-circle diffractometer HEiDi at Heinz Maier- Leibnitz Zentrum (MLZ), Garching (Germany).\cite{Meven_15} A Ge (3 1 1) monochromator was chosen to produce a monochromatic neutron beam with the wavelength of 1.17 Å, and two Er filters were inserted to minimize the $\lambda$/2 and $\lambda$/3 contaminations, respectively. The crystal was mounted on a thin aluminum holder with a small amount of GE varnish and put inside a standard closed-cycle cryostat. The diffracted neutron beam was collected with a $^{3}$He single detector. The integrated intensities of nuclear and magnetic reflections were collected at 2.5 K, 50 K, and 95 K. Rocking-curve scans and $\theta$-$2\theta$ scans were performed for $2\theta<60\text{°}$ and $2\theta>60\text{°}$, respectively. The obtained data sets were normalized to the monitor and corrected by the Lorentz factor. For the dc magnetic suscepetibility ($\chi$), the same piece of crystal was measured in an applied magnetic
field of 1 T parallel to the $\mathit{ab}$ plane and $\mathit{c}$ axis, respectively, using a Quantum Design magnetic property measurement system (MPMS).

\section{Results}

Two-dimensional mesh scans in the ($\mathit{H}$, $\mathit{K}$, 0) reciprocal plane were performed at 2.5 K in the low-$\mathit{Q}$ region to visuallize the magnetic reflections in the ground state. As shown in Figure 1, in addition to the strong nuclear reflections at (1, 0, 0), (0, 1, 0) and (1, 1, 0), additional magnetic reflections appear with the propagation vector $\mathit{k}$ = $\pm$ (1/3, 1/3, 0), consistent with previous observations from neutron powder diffraction and triple-axis single-crystal neutron diffraction on the same compound.\cite{Swainson_10,Plumb_08} No magnetic reflections associated with $\mathit{k}$ = (1/3, -1/3, 0) and $\mathit{k}$ = (-1/3, 1/3, 0) were observed.

\begin{figure}
\centering{}\includegraphics{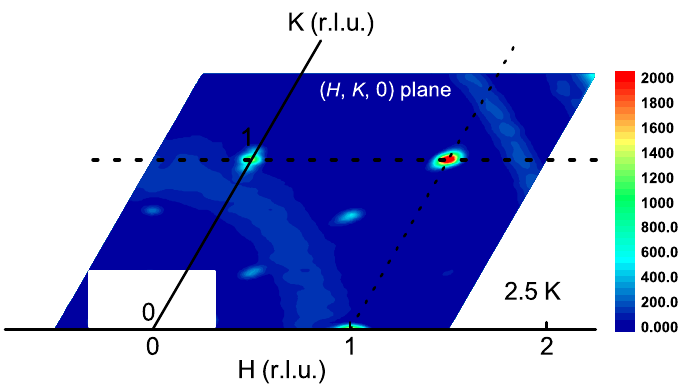}
\caption{The low-$\mathit{Q}$ contour map of the ($\mathit{H}$, $\mathit{K}$, 0) reciprocal plane at 2.5 K. The signals in the white rectangular region are blocked due to the effect of direct neutron beam, and the diffusive rings are due to the diffraction from the aluminum sample holder.}
\end{figure}

\begin{figure}
\centering{}\includegraphics{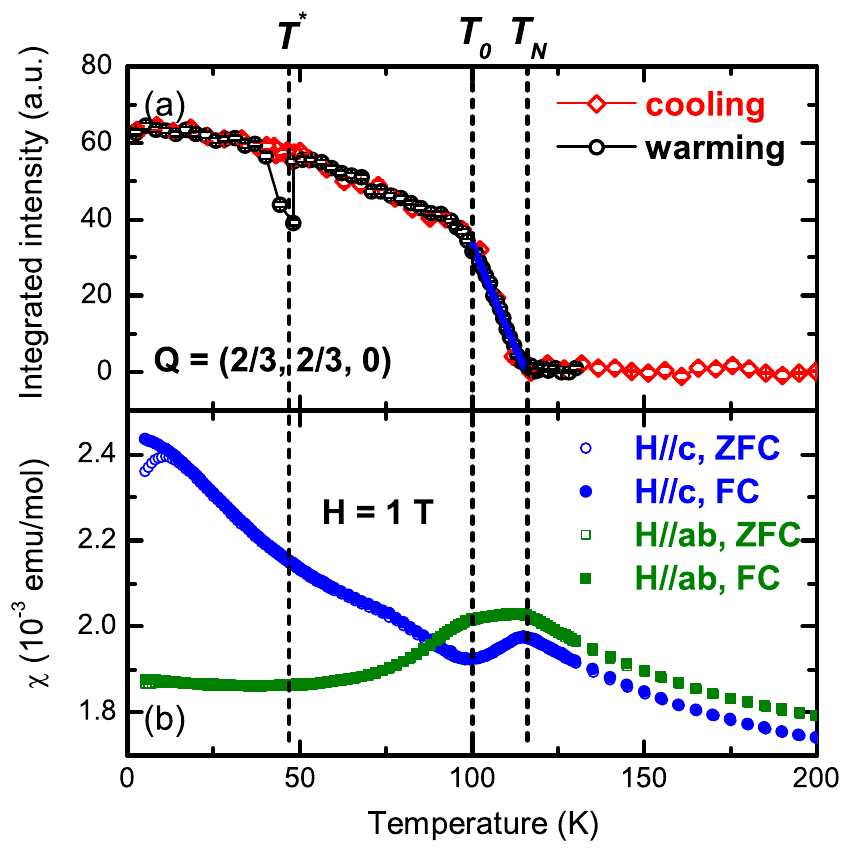}
\caption{(a) The temperature dependence of the integrated intensity of the (2/3, 2/3, 0) magnetic reflection measured during the cooling (red) and warming (black) processes, respectively. The blue solid line represents a fitting to the critical region using the power law. (b) The magnetic susceptibility ($\chi$) of FeCrAs single crystal measured in an applied magnetic field of 1 T parallel to the $\mathit{ab}$ plane (green) and $\mathit{c}$ axis (blue), respectively. The open and filled symbols correspond to the zero-field-cooling (ZFC) and field-cooling (FC) processes, respectively. The three vertical dashed lines mark the two magnetic transitions at $\mathit{\mathit{T_{N}}}$ and $\mathit{T_{O}}$, and the anomaly at $\mathit{T^{*}}$.}
\end{figure}

Fig. 2(a) shows the temperature dependence of the integrated intensity of the (2/3, 2/3, 0) magnetic reflection measured during the cooling and warming processes, respectively. The onset temperature of the antiferromagnetic ordering is determined to be $\mathit{T_{N}}$ =
116(1) K, consistent with the temperature where both $\chi_{c}$ and $\chi_{ab}$ show a peak (as shown in Fig. 2(b)). This is somewhat lower than $\mathit{T_{N}}$ = 125 K as reported in Ref. \onlinecite{Wu_09}, as the value of $\mathit{T_{N}}$ in FeCrAs is known to be sample dependent and can vary between 100 and 125 K depending on the synthesis conditions and sample quality.\cite{Wu_11} Although we were not able to observe the magnetic diffuse scattering due to spin fluctuations above $T_{N}$ due to the small size of the crystal, the critical exponent $\beta$ can be extracted to be $\beta$ = 0.53(4), by fitting the integrated intensities in the critical region (from 100 K to 115 K) using the power law $\mathit{I}$ $\propto$ $\mathit{M^{2}}$ $\propto$ ($\frac{T_{N}-T}{T_{N}})^{2\beta}$ (blue solid line in Fig. 2(a)). It is in excellent agreement with 0.54(5) determined from triple-axis single-crystal neutron diffraction on the same compound,\cite{Plumb_08} suggesting a mean-field type critical behavior in FeCrAs.

Interestingly, the slope of the magnetic order parameter undergoes a clear change around $\mathit{T_{0}}$ \textasciitilde{} 100 K, which coincidences with the dip in $\chi_{c}(T)$ and the bump in $\chi_{ab}(T)$. As explained in detail below, we atrribute $\mathit{T_{0}}$ to the onset of the spin-reorientation of Cr moments from $\mathit{c}$ axis towards the $\mathit{ab}$ plane. Furtheremore, the integrated intensity of the (2/3, 2/3, 0) magnetic reflection shows a sudden drop in between 40 K to 50 K upon warming up (as marked by $\mathit{T^{*}}$), but changes smoothly on cooling. As this anomaly at $\mathit{T^{*}}$ strongly depends on the thermal history and shifts slightly at different thermal cycles (data not shown), it might be due to some magnetic domain behaviors and beyond the scope of this manuscript.

To determine the detailed ground-state magnetic structure of FeCrAs, the integrated intensities of 688 nuclear reflections and 219 magnetic reflections were collected at 2.5 K. The weak reflections with the integrated intensity smaller than twice of the error bar were discarded.
After data reductions, the equivalent nuclear reflections were merged into the unique ones based on the hexagonal symmetry (space group $P\bar{6}2m$), and refined using the FULLPROF program suite.\cite{Rodriguez-Carvajal_93} The extinction correction was performed by refining the single parameter in the isotropic extinction model. No absorption correction was performed, as the neutron absorption coefficients of all three elements are quite small. The refined nuclear structure, as shown in Fig. 3 and Table I, fits the observed integrated intensities very well ($\mathit{R_{F^{2}}}$ = 8.38, $\mathit{R_{F}}$ = 5.03). The scale factor derived from the nuclear structure refinement was then fixed to determine the magnitude of magnetic moments from the refinement of magnetic reflections.

\begin{figure}
\centering{}\includegraphics{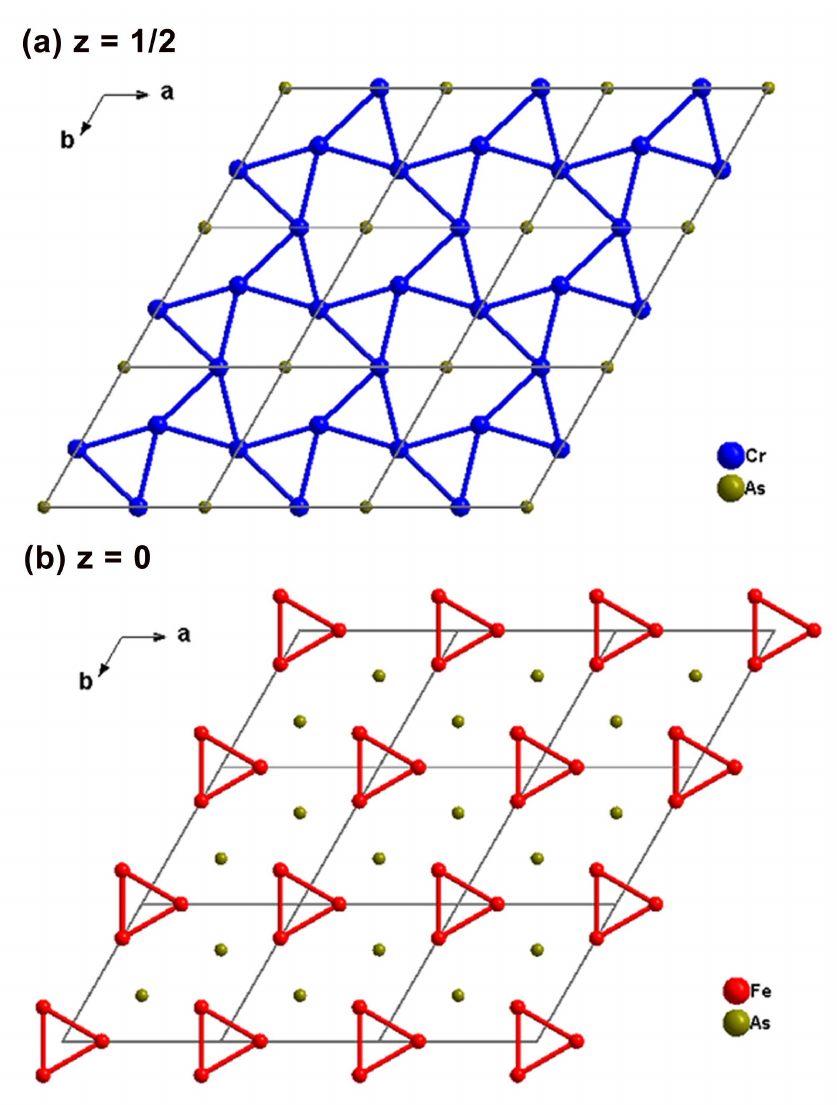}
\caption{The nuclear structure of FeCrAs at 2.5 K projected to the $\mathit{ab}$ plane. The Fe ions form a triangular lattice of trimers, while the Cr ions form a distorted Kagome framework within the basal plane. These planes of Cr Kagome framework (a) and Fe trimers (b) stack alternately along the $\mathit{c}$-axis, with the As ions interspersed throughout both layers.}
\end{figure}

\begin{table}
\caption{Parameters of the nuclear structure of FeCrAs at 2.5 K obtained from refinements of single-crystal neutron diffraction data. {[}Space group: $P\bar{6}2m$, $\mathit{a}$ = $\mathit{b}$ = 6.062(14) Å, $\mathit{c}$ = 3.688(1) Å, $\alpha$ = $\beta$ = 90°, $\gamma$ = 120°.{]}}

\begin{ruledtabular} %
\begin{tabular}{ccccc}
Atom/site & $\mathit{x}$ & $\mathit{y}$ & $\mathsf{\mathit{z}}$ & $B\,$(Å\textsuperscript{2})\tabularnewline
\hline 
Cr ($3g$) & 0.5844(10) & 0 & 1/2 & 0.42(8)\tabularnewline
Fe ($3f$) & 0.2470(4) & 0 & 0 & 0.64(3)\tabularnewline
As ($2c$) & 1/3 & 2/3 & 0 & 0.57(5)\tabularnewline
As ($1b$) & 0 & 0 & 1/2 & 0.57(5)\tabularnewline
\end{tabular}\end{ruledtabular}
\end{table}

Magnetic symmetry analysis was performed using the BasIreps program integrated in the FULLPROF suite. For the space group $P\bar{6}2m$ and $\mathit{k}$ = (1/3, 1/3, 0), the magnetic representation $\Gamma_{mag}$ for the Cr (3$\mathit{g}$) site and the Fe (3$\mathit{f}$) site is the same and can be decomposed as the sum of five irreducible representations (IRs), whose basis vectors are listed in Table II.
\[
\Gamma_{\mathsf{mag}}=\Gamma_{2}^{1}\oplus\Gamma_{3}^{1}\oplus\Gamma_{4}^{1}\oplus\Gamma_{5}^{2}\oplus2\Gamma_{6}^{2}.
\]

\begin{table}[b]
\caption{Basis vectors of the IRs for Cr and Fe atoms in FeCrAs with space group $P\bar{6}2m$ and $\mathit{k}$ = (1/3, 1/3, 0) obtained from representation analysis. The Cr and Fe atoms in one chemical unit cell are defined as Cr(1): (0.5845, 0, 0.5), Cr(2): (0, 0.5845, 0.5),
Cr(3): (0.4155, 0.4155, 0.5), Fe(1): (0.2476, 0, 0), Fe(2): (0, 0.2476, 0), Fe(3): (0.7524, 0.7524, 0), respectively.}

\begin{ruledtabular} %
\begin{tabular}{cccccc}
IRs & $\mathit{\psi_{\nu}}$ & components & Cr(1) & Cr(2) & Cr(3)\tabularnewline
\hline 
$\Gamma_{2}$ & $\mathit{\psi}_{1}$ & Real & (2, 0, 0) & (0, 2, 0) & (1, 1, 0)\tabularnewline
 &  & Imaginary & (0, 0, 0) & (0, 0, 0) & (-$\sqrt{3}$, -$\sqrt{3}$, 0)\tabularnewline
\hline 
$\Gamma_{3}$ & $\mathit{\psi_{\mathsf{2}}}$ & Real & (0, 0, 2) & (0, 0, 2) & (0, 0, -1)\tabularnewline
 &  & Imaginary & (0, 0, 0) & (0, 0, 0) & (0, 0, $\sqrt{3}$)\tabularnewline
\hline 
$\Gamma_{4}$ & $\mathit{\psi_{\mathsf{3}}}$ & Real & (2, 4, 0) & (-4, -2, 0) & (-1, 1, 0)\tabularnewline
 &  & Imaginary & (0, 0, 0) & (0, 0, 0) & ($\sqrt{3}$, -$\sqrt{3}$, 0)\tabularnewline
\hline 
$\Gamma_{5}$ & $\mathit{\psi_{\mathsf{4}}}$ & Real & (0, 0, 2) & (0, 0, -1) & (0, 0, -1)\tabularnewline
 &  & Imaginary & (0, 0, 0) & (0, 0, -$\sqrt{3}$) & (0, 0, -$\sqrt{3}$)\tabularnewline
 & $\mathit{\psi_{\mathsf{5}}}$ & Real & (0, 0, 1) & (0, 0, -2) & (0, 0, 1)\tabularnewline
 &  & Imaginary & (0, 0, $\sqrt{3}$) & (0, 0, 0) & (0, 0, $\sqrt{3}$)\tabularnewline
\hline 
$\Gamma_{6}$ & $\mathit{\psi_{\mathsf{6}}}$ & Real & (2, 0, 0) & (0, -1, 0) & (1, 1, 0)\tabularnewline
 &  & Imaginary & (0, 0, 0) & (0, -$\sqrt{3}$, 0) & ($\sqrt{3}$, $\sqrt{3},$ 0)\tabularnewline
 & $\mathit{\psi_{\mathsf{7}}}$ & Real & (0, 2, 0) & (1, 1, 0) & (-1, 0, 0)\tabularnewline
 &  & Imaginary & (0, 0, 0) & ($\sqrt{3}$, $\sqrt{3}$, 0) & (-$\sqrt{3}$, 0, 0)\tabularnewline
 & $\mathit{\psi_{\mathsf{8}}}$ & Real & (-1, 0, 0) & (0, 2, 0) & (1, 1, 0)\tabularnewline
 &  & Imaginary & (-$\sqrt{3}$, 0, 0) & (0, 0, 0) & ($\sqrt{3}$, $\sqrt{3}$, 0)\tabularnewline
 & $\mathit{\psi_{\mathsf{9}}}$ & Real & (1, 1, 0) & (2, 0, 0) & (0, -1, 0)\tabularnewline
 &  & Imaginary & ($\sqrt{3}$, $\sqrt{3}$, 0) & (0, 0, 0) & (0, -$\sqrt{3}$, 0)\tabularnewline
\end{tabular}\end{ruledtabular}
\end{table}

\begin{figure*}
\centering{}\includegraphics[width=1\textwidth]{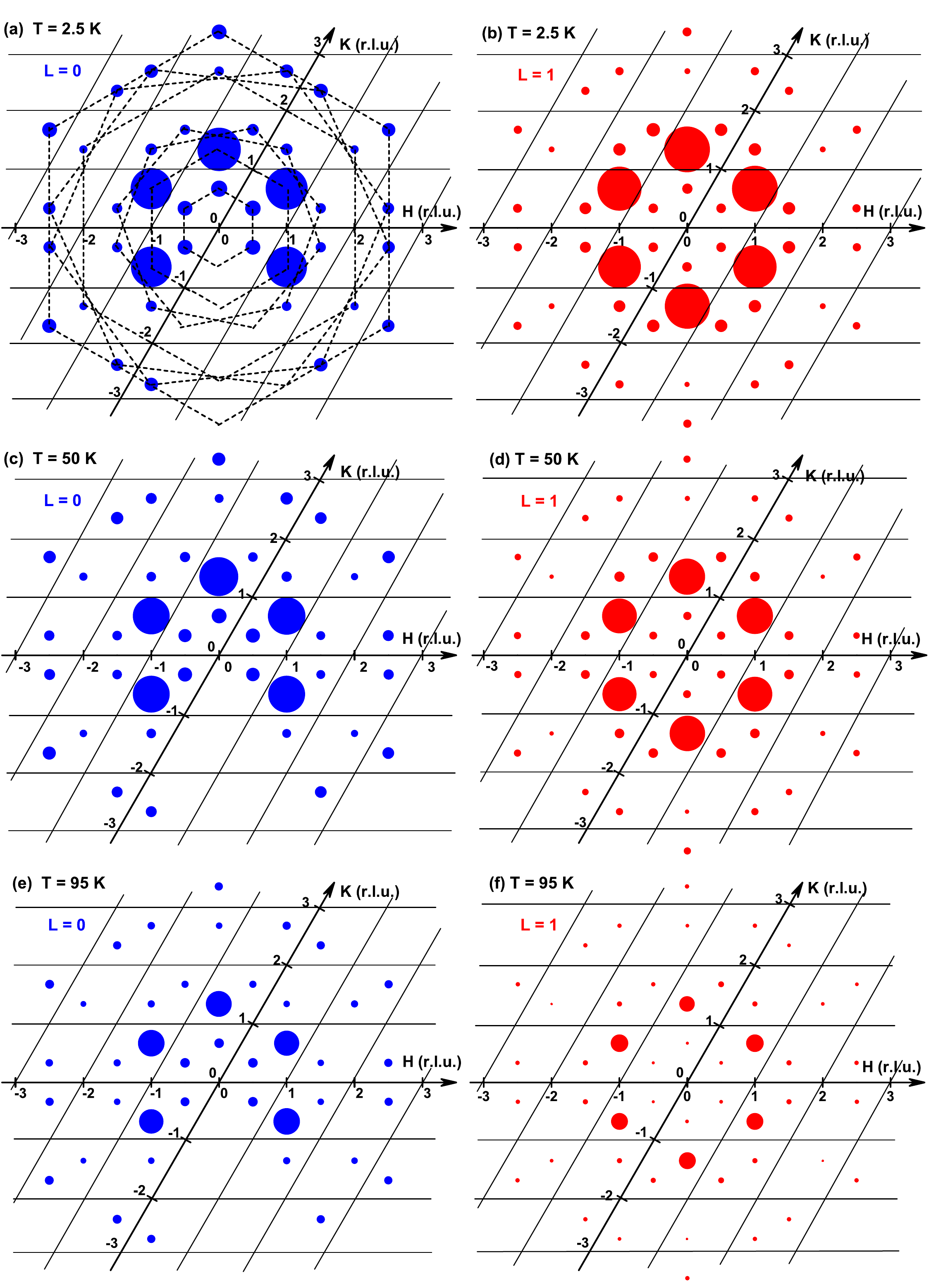}
\caption{The distribution of magnetic diffraction intensities in the ($\mathit{H}$, $\mathit{K}$, 0) and ($\mathit{H}$, $\mathit{K}$, 1) reciprocal planes at 2.5 K (a and b), at 50 K (c and d) and at 95 K (e and f). The sizes of the circles represent the intensities of the magnetic
reflections. The dashed lines in (a) mark the hexagons in which the reflections at their vertices show almost the same insensities within the experimental uncertainty.}
\end{figure*}

The intensity distribution of strong magnetic reflections in the ($\mathit{H}$, $\mathit{K}$, 0) and ($\mathit{H}$, $\mathit{K}$, 1) reciprocal planes at 2.5 K is plotted in Fig. 4(a) and 4(b), respectively, from which a six-fold rotational symmetry is shown. All the observed magnetic reflections are consistent with the propagation vector of $\mathit{k}$ = \textpm{} (1/3, 1/3, 0). The reflections at the vertices of the same hexagons, as marked in Fig. 4(a), show almost the same intensities within the experimental uncertainty. Such a six-fold symmetry of the magnetic diffraction intensities can arise from either a single-domain high-symmetry magnetic configuration with 120° rotational symmetry, or a low-symmetry magnetic configuration residing in three equally-populated domains with the in-plane moments rotated by 120° with respect to each other. Based on neutron powder diffraction, Ref. \onlinecite{Swainson_10} claimed that the ground-state magnetic structure of FeCrAs can be described using a single basis vector. In their model, the moment directions of Cr(1), Cr(2) and Cr(3) are along (1 0 0), (0 -1 0) and (-1 -1 0), with the moment sizes of 4 $\times$ 0.685 $\mu_{B}$, 2 $\times$ 0.685 $\mu_{B}$ and 0.685 $\mu_{B}$, respectively. However, our magnetic symmetry analysis does not generate such a basis vector with the ratio of moment amplitudes between different Cr sites being 4:2:1. As shown below, we find an in-plane spiral antiferromagnetic ground state with different trimers of Cr spins possessing different moment sizes, exhibiting a three-fold rotational symmetry in a single magnetic domain.

\begin{figure}
\centering{}\includegraphics{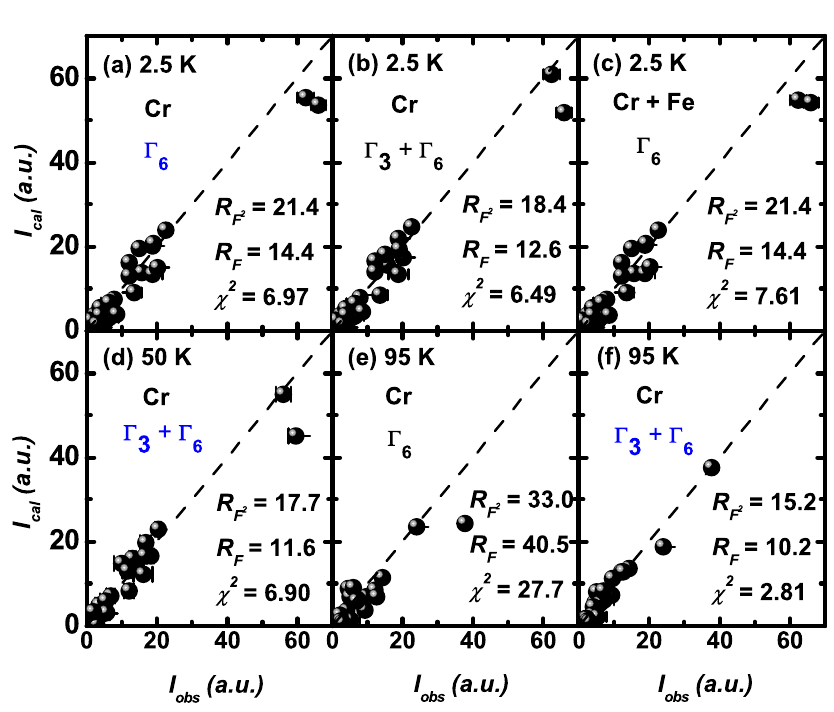}
\caption{Comparisons between the observed and calculated intensities of the magnetic reflections at 2.5 K (a-c), 50 K (d), and 95 K (e, f), respectively. The Fe sublattice is assumed to be non-magnetic in (a), (b), (d), (e) and (f), and magnetic in (c). The out-of-plane moment components described by $\Gamma_{3}$ are included in (b), (d) and (f), while purely in-plane moments described by $\Gamma_{6}$ are considered in (a), (c) and (e). The blue color marks the adopted model at different temperature.}
\end{figure}

Due to the six-fold rotational symmetry of the magnetic intensities in ($\mathit{H}$, $\mathit{K}$, 0), ($\mathit{H}$, $\mathit{K}$, 1) and ($\mathit{H}$, $\mathit{K}$, 2) (data not shown) planes, every six magnetic reflections from the same hexagon as shown in Fig.
4(a) were averaged into one unique reflection for simplicity and tested by different magnetic structure models. The integrated intensities of magnetic reflections from our single crystal do not support any magnetic structure described by a single basis vector as listed in Table II. Instead, a combination of different basis vectors have to be implemented in our refinement. As a result, a high-symmetry magnetic configuration with 120° rotational symmetry described by the irreducible presentation of $\Gamma_{6}$ with the combination of two basis vectors,
$\psi_{8}$ and $\psi_{9}$, yields a good fitting to the observed magnetic intensities at 2.5 K ($\mathit{R_{F^{2}}}$ = 18.4, $\mathit{R_{F}}$ = 12.6, $\chi^{2}$ = 6.49), as shown in Fig. 5(a). In this model, the Cr moments order spirally in the $\boldsymbol{ab}$ plane with
their amplitudes fluctuating in the form of a spin density wave (see Fig. 6(a)), showing a 3-fold rotational symmetry in a single magnetic domain. Adding an out-of-plane moment component described by $\Gamma_{3}$ (basis vector $\psi_{2}$) into this spiral configuration can slightly improve the fitting (Fig. 5(b)), while destroying the overall 3-fold rotational symmetry in the meantime. Therefore, the spiral ordering of the Cr moments purely in the $\boldsymbol{ab}$ plane is believed to be the ground-state magnetic structure of FeCrAs, owing to the
higher symmetry it exhibits. However, the magnetic reflections collected at 95 K are clearly better fitted with the irreducible representation of $\Gamma_{3}$ + $\Gamma_{6}$, as Fig. 5(e) and (f) show, suggesting large out-of-plane components of the Cr moments. At 50 K, we also use the mixture of $\Gamma_{3}$ and $\Gamma_{6}$ to describe the intermediate-temperature magnetic structure (Fig. 5(d)), as the spin canting is believed to occur gradually with increasing temperature. For all three temperatures, inclusion of the magnetic moments on the
Fe sublattice with the same magnetic basis vectors as Cr does not improve the refinements, as shown in Fig. 5(c). The upper limit of the Fe moment is estimated to be less than 0.1 $\mu_{B}$, consistent with 0.1 $\pm$ 0.03 $\mu_{B}$ determined from M\"ossbauer spectroscopy study.\cite{Footnote} Since including the magnetic moments on the Fe sites does not change the magnetic structure of Cr sublattice, we will only present the results assuming a non-magnetic Fe sublattice in the discussions below.

\begin{figure*}
\centering{}\includegraphics[width=1\textwidth]{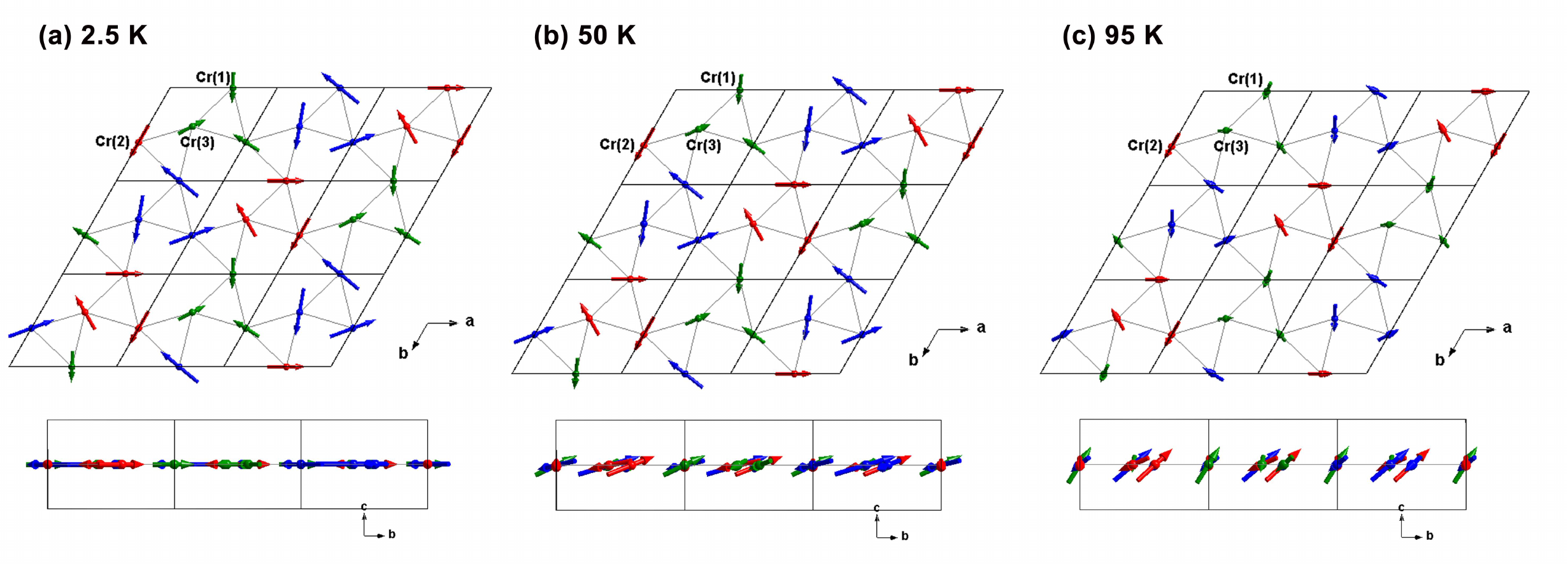}
\caption{The refined magnetic structure of FeCrAs in one magnetic domain at 2.5 K (a) and 95 K (b), viewed along the $\mathit{c}$ axis (top) and $\mathit{a}$ axis (bottom), respectively.}
\end{figure*}

\begin{table*}
\caption{Parameters about the magnetic structure of Cr sublattice in FeCrAs at 2.5 K, 50 K, and 95 K, respectively. The Cr atoms in one chemical unit cell are defined as Cr(1): (0.5845, 0, 0.5), Cr(2): (0, 0.5845, 0.5), and Cr(3): (0.4155, 0.4155, 0.5), respectively. The numbers
in the parentheses correspond to the error bars of the calculated total moment sizes.}

\begin{ruledtabular} %
\begin{tabular}{cc|c||c||cc|cc|cc}
\multirow{2}{*}{Atom sites} & \multirow{2}{*}{Translation} & \multicolumn{4}{c|}{2.5 K} & \multicolumn{2}{c|}{50 K} & \multicolumn{2}{c}{95 K}\tabularnewline
 &  & \multicolumn{3}{c}{($\mathit{M_{a}}$, $\mathit{M_{b}}$, $\mathit{M_{c}}$) ($\mu_{B})$} & $\mathit{M_{tot}}$($\mu_{B})$ & ($\mathit{M_{a}}$, $\mathit{M_{b}}$, $\mathit{M_{c}}$) ($\mu_{B})$ & $\mathit{M_{tot}}$($\mu_{B})$ & ($\mathit{M_{a}}$, $\mathit{M_{b}}$, $\mathit{M_{c}}$) ($\mu_{B})$ & $\mathit{M_{tot}}$($\mu_{B})$\tabularnewline
\hline 
Cr(1) & (0, 0, 0) & \multicolumn{3}{c}{(0.47, 1.02, 0)} & 0.89(16) & (0.35, 0.87, 0.58) & 0.95(18) & (0.05, 0.37, 0.83) & 0.90(22)\tabularnewline
 & (1, 0, 0)/(0, 1, 0) & \multicolumn{3}{c}{(-1.58, -1.02, 0)} & 1.38(12) & (-1.38, -0.87, -0.29) & 1.24(11) & (-0.69, -0.37, -0.42) & 0.73(17)\tabularnewline
 & (2, 0, 0)/(0, 2, 0) & \multicolumn{3}{c}{(1.11, 0, 0)} & 1.11(14) & (1.03, 0, -0.29) & 1.07(13) & (0.64, 0, -0.42) & 0.76(18)\tabularnewline
\hline 
Cr(2) & (0, 0, 0) & \multicolumn{3}{c}{(0, 1.11, 0)} & 1.11(14) & (0, 1.03, 0.58) & 1.18(17) & (0, 0.64, 0.83) & 1.05(24)\tabularnewline
 & (1, 0, 0)/(0, 1, 0) & \multicolumn{3}{c}{(-1.02, -0.55, 0)} & 0.89(16) & (-0.87, -0.51, -0.29) & 0.81(17) & (-0.37, -0.32, -0.42) & 0.54(22)\tabularnewline
 & (2, 0, 0)/(0, 2, 0) & \multicolumn{3}{c}{(1.02, -0.55, 0)} & 1.38(12) & (0.87, -0.51, -0.29) & 1.24(11) & (0.37, -0.32, -0.42) & 0.73(17)\tabularnewline
\hline 
Cr(3) & (0, 0, 0) & \multicolumn{3}{c}{(0.55, -0.47, 0)} & 1.11(14) & (0.51, -0.35, -0.29) & 0.81(18) & (0.32, -0.05, -0.42) & 0.54(22)\tabularnewline
 & (1, 0, 0)/(0, 1, 0) & \multicolumn{3}{c}{(0.55, 1.58, 0)} & 0.89(16) & (0.51, 1.38, 0.58) & 1.34(16) & (0.32, 0.69, 0.83) & 1.02(23)\tabularnewline
 & (2, 0, 0)/(0, 2, 0) & \multicolumn{3}{c}{(-1.11, -1.11, 0)} & 1.38(12) & (-1.03, -1.03, -0.29) & 1.07(13) & (-0.64, -0.64, -0.42) & 0.76(18)\tabularnewline
\end{tabular}\end{ruledtabular}
\end{table*}

The magnetic parameters of the Cr sublattice at 2.5 K, 50 K, and 95 K obtained through the refinements are listed in Table III. The magnetic propagation vector $\mathit{k}$ = (1/3, 1/3, 0) indicates that the magnetic unit cell is 3$\times$3 times of the chemical unit cell, and the magnetic moment on a specific Cr site gets repeated upon the translation of 3 chemical unit cells along the $\mathit{a}$ and $\mathit{b}$ directions. Therefore, there are 9 distinct Cr sites in total with different moment sizes in a magnetic unit cell, forming the spin-density-wave
type antiferromagnetic order since the net moment is zero. We find an in-plane spiral antiferromagnetic ground state described by $\Gamma_{6}$ ( $\psi_{8}$ + $\psi_{9}$), with different trimers of Cr spins (denoted by different colors in Fig. 6) possessing different
moment sizes, exhibiting a three-fold rotational symmetry. The moment sizes at all Cr sites in Table III depends purely on the coefficients of the two basis vectors, $\psi_{8}$ and $\psi_{9}$, which can be well determined by the refinements of the magnetic reflections. At 2.5 K, the moment sizes of the Cr ions were found to vary between 0.8 and 1.4 $\mu_{B}$ on different sites, which are significantly smaller than the theoretical spin-only magnetic moment of 3.87 $\mu_{B}$ for Cr$^{3+}$ ions ($\mathit{S}$ = 3/2),\cite{Footnote2} which can be due to strong magnetic frustration in the distorted Kagome Cr lattice and/or the itinerant nature of the Cr magnetism..

The magnetic structures of the Cr sublattice in one magnetic domain in FeCrAs at 2.5 K, 50 K, and 95 K were visualized in Fig. 6(a), (b), and (c), respectively. Compared with 2.5 K, the in-plane magnetic structure at 50 K only shows a minor change. However, it is quite clear that the in-plane moments at different Cr sites are significantly suppressed at 95 K, while the moment components along the $\mathit{c}$ direction become quite large in contrast. All Cr moments are gradually canted towards the c axis with increasing temperature. The distinct
temperature-dependent behaviors of the in-plane and out-of-plane Cr moments are better illustrated in Fig. 7(b), in which the in-plane and out-of-plane moment sizes of Cr(1) site were plotted as a function of temperature (the other Cr sites show similar tendencies). Accordingly,
the intensities of strong magnetic reflections with $\mathit{L}$ = 1 decrease more dramatically with increasing temperature (see Fig. 7(a) and Fig. 4), compared with those with $\mathit{L}$ = 0, as the magnetic scattering at these $\mathit{Q}$ points are more sensitive to the suppression of in-plane Cr magnetic moments.

\begin{figure}[t]
\centering{}\includegraphics{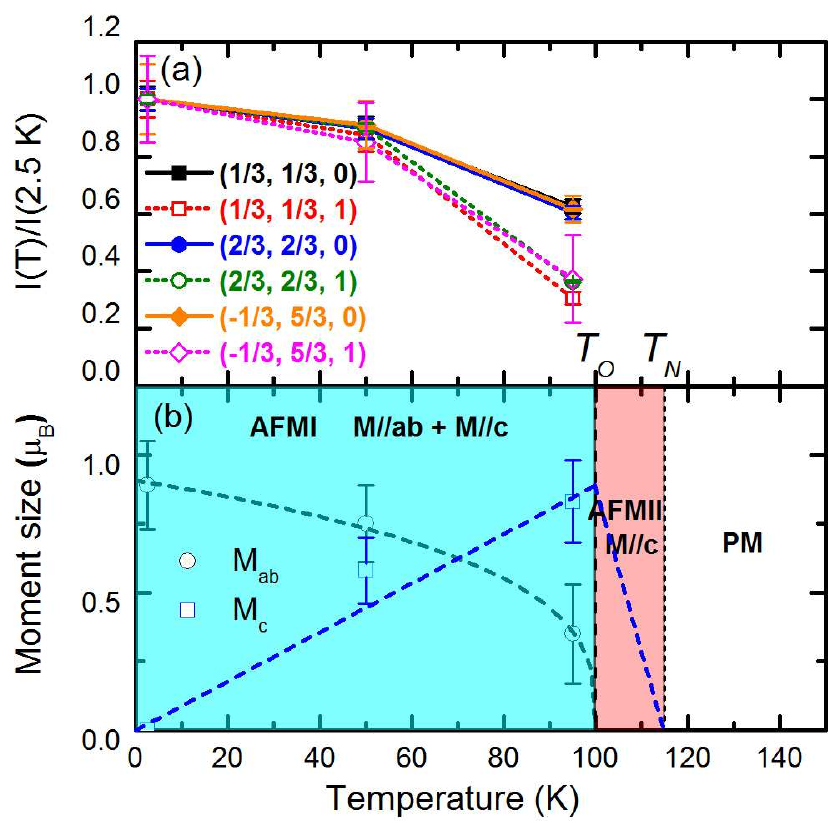}
\caption{The temperature dependencies of the normalized intensities of strong magnetic reflections (a), as well as the in-plane ($\mathit{M_{ab}}$) and out-of-plane ($\mathit{M_{c}}$) moment sizes of Cr(1) site (b) (the other Cr sites show similar tendencies). $\mathit{T_{N}}$ and $\mathit{T_{O}}$ correspond to the antiferromagnetic transition temperature and the onset temperature of the in-plane Cr moments, respectively. The green, red and white regions correspond to the AFMI phase with both in-plane and out-of-plane Cr moments, the AFMII phase with only the out-of-plane Cr moments and the paramagnetic phase, respectively.}
\end{figure}

\section{Discussion and Conclusion}

Since in the paramagnetic state above $\mathit{T_{N}}$ all static moments on different Cr sites are supposed to vanish and there is no additional phase transitions in between $\mathit{T_{O}}$ = 100 K and $\mathit{T_{N}}$ as shown in Fig. 2, it is reasonable to speculate
that the out-of-plane Cr moments reach the maximum sizes at $\mathit{T_{O}}$ and weaken when the temperature approaches $\mathit{T_{N}}$. Taking into account the susceptibility behaviors in Fig. 2, it is very likely that the onset of in-plane Cr moments occurs below $\mathit{T_{O}}$. Unfortunately, due to the extreme weakness of the magnetic reflections above 100 K, we were not able to collect another data sets in between $\mathit{T_{N}}$ and $\mathit{T_{O}}$ to further verify our scenario. However, our picture is consistent with the magnetic susceptibility data in Fig. 2. With the initial magnetic ordering with the Cr moments purely along the $\mathit{c}$ axis, it is expected that $\chi_{c}$ peaks at $\mathit{T_{N}}$ and $\chi_{ab}$ flattens below $\mathit{T_{N}}$. Upon further cooling, $\chi_{c}$ reverses below $\mathit{T_{O}}$ due to the increased contribution from in-plane Cr moments, and $\chi_{ab}$ decreases due to reduced contribution from out-of-plane Cr moments.

Separated by the two magnetic transitions at $\mathit{T_{N}}$ and $\mathit{T_{O}}$, there exists three distinct phases for FeCrAs, including the paramagnetic phase above $\mathit{T_{N}}$, the AFMII phase with only the out-of-plane Cr moments in between $\mathit{T_{N}}$ and $\mathit{T_{O}}$, and the AFMI phase with both in-plane and out-of-plane Cr moments below $\mathit{T_{O}}$, respectively. The spin reorientation in the Cr sublattice from $\mathit{c}$ axis towards the $\mathit{ab}$ plane in FeCrAs resembles the spin-flip transition observed in elemental Cr,\cite{Fawcett_88} in which a spin-flip transition was observed at $\mathit{T_{SF}}$ $\cong$ 123 K from a high-temperature transversely polarized phase (with the moment direction $\mathbf{M}$ perpendicular to the wave vector $\mathbf{Q}$) to a low-temperature longitudinally polarized phase (with $\mathbf{M}$ parallel to $\mathbf{Q}$). In our case, the high-temperture AFMII phase is also transversely polarized (with $\mathbf{M}$ along $\mathit{c}$ and $\mathbf{Q}$ in the $\mathit{ab}$ plane), and the low-temperature AFMI phase is close to longitudinally polarized (with $\mathbf{M}$ canted but close to the $\mathit{ab}$ plane). It is believed that the coupling between SDW and strain is responsible for the spin-flip transition in Cr,\cite{Cowan_78} suggesting that spin-lattice coupling may be important in FeCrAs as reported in Ref. \onlinecite{Akrap_14}. The striking similarity between FeCrAs and the model itinerant magnet Cr in aspects of spin-flip transitions and SDW-type antiferromagnetism nature might suggest the dominance of itinerant magnetism in FeCrAs. This is in line with the "itinerant type'', high-energy spin fluctuations as revealed by recent INS experiment on powder samples of FeCrAs.\cite{Plumb_08} Further INS measurements on single-crystal samples of FeCrAs will be crucial for understanding the nature of its intriguing magnetism.

In summary, the magnetic structure of "nonmetallic metal'' FeCrAs, a compound with the characters of both metals and insulators, were examined as a function of temperature using single-crystal neutron diffraction. In the ground state, the Cr sublattice shows an in-plane spiral antiferromagnetic order with $\mathit{k}$ = (1/3, 1/3, 0). The moment sizes of the Cr ions were found to be small, due to strong magnetic frustration in the distorted Kagome framework or the itinerant nature of the Cr magnetism, and vary between 0.8 and 1.4 $\mu_{B}$ on different sites as expected for a SDW type order. The upper limit of the moment on the Fe sublattice is estimated to be less than 0.1 $\mu_{B}$. With increasing temperature up to 95 K, the Cr moments cant out of the $\mathit{ab}$ plane gradually, with the in-plane components being suppressed and the out-of-plane components increasing in contrast. This spin-reorientation of Cr moments can explain the dip in $\chi_{c}$ and the kink in the magnetic order parameter at $\mathit{T_{O}}$ \textasciitilde{} 100 K. We have also dicussed the similarity between FeCrAs and the model itinerant magnet Cr, which exhibits spin-flip transitions and SDW-type antiferromagnetism.

\bibliographystyle{apsrev} \bibliographystyle{apsrev}
\begin{acknowledgments}
This work is based on experiments performed at the HEiDi instrument operated by Jülich Centre for Neutron Science (JCNS) at the Heinz Maier-Leibnitz Zentrum (MLZ), Garching, Germany. Work at the University of Toronto was supported by the Natural Sciences and Engineering Research Council (NSERC) of Canada through the Collaborative Research and Training Experience (CREATE) program (432242-2013) and Discovery Grant No. RGPIN-2014-06071 and RGPIN-2014-04554. W.T.J. would like to acknowledge S. Gao and B. Yuan for helpful discussions.

\bibliographystyle{apsrev}

\begin{thebibliography}{10}
\bibitem{Footnote2}
The valence state of Cr in FeCrAs is hard to conclude due to the itinerant nature of its magnetism. In our refinements, we were using the magnetic form factors of Cr\textsuperscript{3+} ions for simplicity. Using the magnetic form factors of Cr\textsuperscript{2+} ions yields similar results.
\end{thebibliography}

\end{acknowledgments}

\end{document}